\documentclass[12pt]{article}
\usepackage{graphicx}
\newcommand{\Mgf}{\ensuremath{\mathcal{M}}}

\begin{document}
\title{Critical exponents of plane meanders}
\author{ 
Iwan Jensen\thanks{e-mail: I.Jensen@ms.unimelb.edu.au}\mbox{ }
and Anthony J. Guttmann\thanks{e-mail: tonyg@ms.unimelb.edu.au}, \\
Department of Mathematics \& Statistics, \\
The University of Melbourne,\\
Victoria 3010, Australia}
\maketitle
\bibliographystyle{plain}
\begin{abstract}
Meanders form a set of combinatorial problems concerned with the enumeration 
of self-avoiding loops crossing a line through a given number of points, $n$. 
Meanders are considered distinct up to any smooth deformation leaving the 
line fixed. We use a recently developed algorithm, based on transfer matrix 
methods, to enumerate plane meanders. This allows us to calculate the number 
of closed meanders up to $n=48$, the number of open meanders up to $n=43$,
and the number of semi-meanders up to $n=45$. The analysis of the series
yields accurate estimates of both the critical point and critical exponent, 
and shows that a recent conjecture for the exact value of the semi-meander 
critical exponent is unlikely to be correct, while the conjectured exponent 
value for closed and open meanders is not inconsistent with the results from 
the analysis.
\end{abstract}

Meanders form a set of unsolved combinatorial problems concerned with the 
enumeration of self-avoiding loops crossing a line through a given number 
of points \cite{LZ}. Meanders are considered distinct up to any smooth 
deformation leaving the line fixed. This problem seems to date back at least 
to the work of Poincar\'e on differential geometry \cite{Poincare}. Since 
then it has from time to time been studied by mathematicians in various 
contexts such as the folding of a strip of stamps \cite{Touchard,Koehler} or 
folding of maps \cite{Lunnon}. More  recently it has been related to 
enumerations of ovals in planar algebraic curves \cite{Arnold} and the 
classification of 3-manifolds \cite{KS}. During the last decade or so it has 
received considerable attention in other areas of science. In computer science 
meanders are related to the sorting of Jordan sequences \cite{HMRT} and have 
been used for lower bound arguments \cite{AM}. In physics meanders are 
relevant to the study of compact foldings of polymers \cite{FGG1,FGG2}, 
properties of the Temperley-Lieb algebra \cite{FGG3,Francesco}, 
matrix models \cite{Makeenko,SS}, and models of low-dimensional
gravity \cite{FGG4}. 

\begin{figure}[h]
\includegraphics{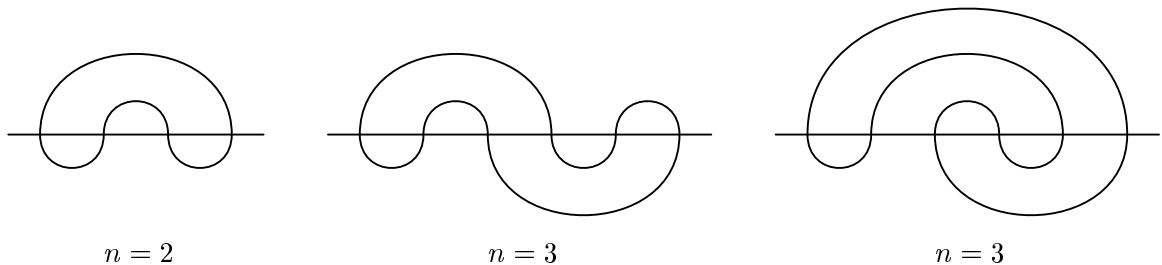}
\caption{\label{fig:meanclose} A few examples of closed meanders of
order 2 and 3, respectively. }
\end{figure}

A {\em closed meander} of order $n$ is a closed self-avoiding loop 
crossing an infinite line $2n$ times (see figure~\ref{fig:meanclose}). 
The meandric number $M_n$ is simply the number of such meanders distinct 
up to smooth transformations. Note that each meander forms a single 
connected loop. The number of closed meanders is expected to grow 
exponentially, with a sub-dominant term given by a critical exponent,
$M_n \sim  C R^{2n}/n^{\alpha}$. The exponential growth constant $R$
is often called the {\em connective constant}.
Thus the generating function  is expected to behave as 

\begin{equation}\label{eq:meangen}
\Mgf(x) = \sum_{n=1}^{\infty} M_n x^n \sim A(x)(1 - R^2 x)^{\alpha-1},
\end{equation}
and hence have
a singularity at $x_c=1/R^2$ with exponent $\alpha -1$. 
The first meandric numbers are $M_1 =1$, $M_2 =2$ and  $M_3 =8$.
One can extend the definition to {\em multi-component systems of 
closed meanders}, where we allow configurations with several 
disconnected closed loops.
The meandric numbers $M_n^{(k)}$ are then the number of meanders with
$2n$ crossings and $k$ independent loops.

An {\em open meander} of order $n$ is a self-avoiding curve running from 
west to east while crossing an infinite line $n$ times. The number
of such curves is $m_n$ and we can define a generating function for
this problem in analogy with (\ref{eq:meangen}). It should be noted
\cite{LZ} that $M_n = m_{2n-1}$, and hence the critical exponent is 
identical to that of closed meanders and the connective constant is $R$.

\begin{figure}[h]
\includegraphics{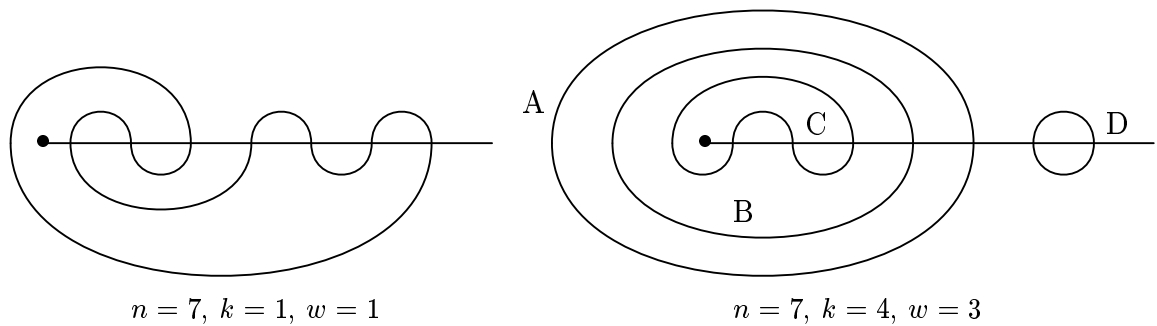}
\caption{\label{fig:meansemi} Two examples of semi-meanders. The first
one has a single loop, wind around the origin once, and contain
7 crossings. The second semi-meander has 4 loops (labelled A--D),
wind around the origin 3 times, and again contain 7 crossings.
}
\end{figure}

Finally, instead of looking at intersections with an infinite line
one could consider a semi-infinite line and allow the curve to wind
around the end-point of the line \cite{FGG1}. A {\em semi-meander} of 
order $n$ is a closed self-avoiding loop crossing the semi-infinite 
line $n$ times. The number of semi-meanders of order $n$ is denoted by 
$\overline{M}_n\sim C' \overline{R}/n^{\overline{\alpha}}$ 
and we define a generating function as in (\ref{eq:meangen}).
In this case a further interesting generalization is to study the number
of semi-meanders $\overline{M}_n(w)$ which wind around the  end-point of 
the line exactly $w$ times. Again we could also study systems of 
multi-component semi-meanders according to the number of independent loops.
Two semi-meanders are shown in figure~\ref{fig:meansemi}.

In a recent paper it was argued that the meander problem is
related to the gravitational version of a certain loop model \cite{FGG4}.
From the conformal field theory of the model, conjectures were proposed 
for the exact critical exponent of closed and open meanders,
$\alpha= (29+\sqrt{145})/12 = 3.4201328\ldots$, 
as well as the exponent for semi-meanders,
$\overline{\alpha}= 1+\sqrt{11}(\sqrt{29}+\sqrt{5})/24 = 2.0531987\ldots$.
This work has recently been extended to multi-component systems of 
closed and semi-meanders \cite{FGJ2000}. Conjectures were then given
for the critical exponents as functions of the loop-fugacity $q$. 
These were checked numerically \cite{FGJ2000} and found to 
be correct within numerical error. In this
Letter we analyse extended series for the meander generating functions.
Using differential approximants we obtain accurate estimates for
the exponents and find that the conjecture for the semi-meander
exponent is unlikely to be correct, while the conjecture
for closed meanders is not inconsistent with the results
from the analysis.

The difficulty in the enumeration of most interesting combinatorial problems
is that, computationally, they are of exponential complexity. Initial efforts 
at computer enumeration of meanders were based on direct counting. 
Lando and Zvonkin \cite{LZ} studied closed meanders, open meanders
and multi-component systems of closed meanders, while 
Di Francesco {\em et al.} \cite{FGG2} studied semi-meanders. In this Letter 
we use a new and improved algorithm \cite{Jensen99}, based on transfer 
matrix methods, to enumerate various meander problems such as closed, 
open and semi-meanders. The method is similar to the 
transfer matrix technique devised by Enting \cite{Ent} in his pioneering 
work on the enumeration of self-avoiding polygons. The first terms in the 
series for the meander generating function can be calculated using transfer 
matrix techniques.  This involves drawing an intersection  
perpendicular to the infinite line. Meanders are enumerated by successive
moves of the intersection, so that one crossing at a time is added to the
meanders. A preliminary description of the algorithm can be found
in \cite{Jensen99} and further details will appear elsewhere.
A very closely related algorithm was used and described in \cite{FGJ2000}.

\begin{table}
\caption{\label{tab:closed} The number, $M_n$, of connected closed 
meanders with $2n$ crossings.}
\begin{center}
\small
\begin{tabular}{rrrrrr} \hline \hline
$n$ & \multicolumn{1}{c}{$M_n$} &
$n$ & \multicolumn{1}{c}{$M_n$} & 
$n$ & \multicolumn{1}{c}{$M_n$} \\ \hline 
1& 1 &     9  & 933458 &         17 & 59923200729046  \\                 
2& 2 &     10 & 8152860 &        18 & 608188709574124 \\                  
3& 8 &     11 & 73424650 &       19 & 6234277838531806 \\                 
4& 42&     12 & 678390116 &      20 & 64477712119584604 \\                 
5& 262&    13 & 6405031050 &     21 & 672265814872772972 \\              
6& 1828 &  14 & 61606881612 &    22 & 7060941974458061392 \\              
7& 13820 & 15 & 602188541928 &   23 & 74661728661167809752 \\              
8& 110954 &16 & 5969806669034 &  24 & 794337831754564188184 \\            
\hline \hline
\end{tabular}
\end{center}
\end{table}

\begin{table}
\caption{\label{tab:open} The number, $m_n$, of connected open
meanders with $n$ crossings.}
\begin{center}
\small
\begin{tabular}{rrrrrr} \hline \hline
$n$ & \multicolumn{1}{c}{$m_n$} &
$n$ & \multicolumn{1}{c}{$m_n$} & 
$n$ & \multicolumn{1}{c}{$m_n$} \\ \hline 
1 & 1 & 16 & 252939 & 31 & 5969806669034 \\
2 & 1 & 17 & 933458 & 32 & 15012865733351 \\
3 & 2 & 18 & 2172830 & 33 & 59923200729046 \\
4 & 3 & 19 & 8152860 & 34 & 151622652413194 \\
5 & 8 & 20 & 19304190 & 35 & 608188709574124 \\
6 & 14 & 21 & 73424650 & 36 & 1547365078534578 \\
7 & 42 & 22 & 176343390 & 37 & 6234277838531806 \\
8 & 81 & 23 & 678390116 & 38 & 15939972379349178 \\
9 & 262 & 24 & 1649008456 & 39 & 64477712119584604 \\
10 & 538 & 25 & 6405031050 & 40 & 165597452660771610 \\
11 & 1828 & 26 & 15730575554 & 41 & 672265814872772972 \\
12 & 3926 & 27 & 61606881612 & 42 & 1733609081727968492 \\
13 & 13820 & 28 & 152663683494 & 43 & 7060941974458059344 \\
14 & 30694 & 29 & 602188541928 &  &  \\
15 & 110954 & 30 & 1503962954930 &  &  \\
\hline \hline
\end{tabular}
\end{center}
\end{table}

\begin{table}
\caption{\label{tab:semi} The number, $\overline{M}_n$, of connected 
semi-meanders with $n$ crossings.}
\begin{center}
\small
\begin{tabular}{rrrrrr} \hline \hline
$n$ & \multicolumn{1}{c}{$\overline{M}_n$} &
$n$ & \multicolumn{1}{c}{$\overline{M}_n$} & 
$n$ & \multicolumn{1}{c}{$\overline{M}_n$} \\ \hline 
1 & 1 & 16 & 1053874 & 31 & 42126805350798 \\
2 & 1 & 17 & 3328188 & 32 & 137494070309894 \\
3 & 2 & 18 & 10274466 & 33 & 455792943581400 \\
4 & 4 & 19 & 32786630 & 34 & 1493892615824866 \\
5 & 10 & 20 & 102511418 & 35 & 4967158911871358 \\
6 & 24 & 21 & 329903058 & 36 & 16341143303881194 \\
7 & 66 & 22 & 1042277722 & 37 & 54480174340453578 \\
8 & 174 & 23 & 3377919260 & 38 & 179830726231355326 \\
9 & 504 & 24 & 10765024432 & 39 & 600994488311709056 \\
10 & 1406 & 25 & 35095839848 & 40 & 1989761816656666392 \\
11 & 4210 & 26 & 112670468128 & 41 & 6664356253639465480 \\
12 & 12198 & 27 & 369192702554 & 42 & 22124273546267785420 \\
13 & 37378 & 28 & 1192724674590 & 43 & 74248957195109578520 \\
14 & 111278 & 29 & 3925446804750 & 44 & 247100408917982623532 \\
15 & 346846 & 30 & 12750985286162 & 45 & 830776205506531894760 \\
\hline \hline
\end{tabular}
\end{center}
\end{table}

The enumerations undertaken thus far are too numerous to detail here. We 
only give the results for connected closed meanders $M_n$, open meanders
$m_n$, and semi-meanders which wind around the origin any
number of times and have only a single loop $\overline{M}_n$. The numbers 
of such meanders are listed in Table~\ref{tab:closed}-Table~\ref{tab:semi}. 

\begin{table}
\caption{\label{tab:analysis} Estimates of the critical points and
exponents of the meander generating functions for closed, open and 
semi-meanders, as obtained from 2. order differential approximants.
$L$ is the degree or the inhomogeneous polynomial.}
\begin{center}
\small
\begin{tabular}{rllllll}
\hline \hline
 & \multicolumn{2}{c}{Closed meanders} & 
 \multicolumn{2}{c}{Open meanders} & 
 \multicolumn{2}{c}{Semi-meanders} \\
\hline 
\multicolumn{1}{c}{L} &
\multicolumn{1}{c}{$x_c$} &
\multicolumn{1}{c}{$\alpha-1$} &
\multicolumn{1}{c}{$x_c$} &
\multicolumn{1}{c}{$\alpha-1$} &
\multicolumn{1}{c}{$x_c$} &
\multicolumn{1}{c}{$\overline{\alpha}-1$} \\
\hline 
0  & 0.08154671(24)& 2.42104(42)
   & 0.28556361(40)& 2.42129(36)                         
   & 0.285564437(10)& 1.053693(12)\\
1  & 0.08154684(14)& 2.42084(33) 
   & 0.28556416(19)& 2.42075(20)  
   & 0.28556448(10)& 1.05362(16) \\
2  & 0.081546912(59)& 2.42079(45) 
   & 0.28556418(64) & 2.42072(63)  
   & 0.28556447(13)& 1.05358(27) \\                                           
3  & 0.081546916(84)& 2.42069(17)
   & 0.28556386(33)& 2.42109(34)
   & 0.285564436(31)& 1.053693(47) \\
4  & 0.081546950(46)& 2.42061(10)                              
   & 0.28556390(50)& 2.42101(46) 
   & 0.285564433(29)& 1.053700(34)\\
5  & 0.081546901(82)& 2.42074(18) 
   & 0.28556406(10)& 2.42088(13) 
   & 0.285564437(24)& 1.053692(32) \\
6  & 0.081546917(67)& 2.42065(21)  
   & 0.28556394(28)& 2.42101(32) 
   & 0.285564425(65)& 1.053699(96)\\
7  & 0.081546910(72)& 2.42070(21) 
   & 0.28556407(10)& 2.42088(12)  
   & 0.285564413(58)& 1.053717(71)\\
8  & 0.08154682(16)& 2.42090(30)  
   & 0.28556408(10)& 2.42087(13) 
   & 0.285564425(46)& 1.053706(54)\\
9  & 0.08154668(32)& 2.42115(60) 
   & 0.285564096(83)& 2.42083(11) 
  & 0.285564434(52)& 1.053692(79)\\
10 & 0.08154671(26)& 2.42107(45)  
   & 0.28556414(16)& 2.42078(18) 
   & 0.285564433(47)& 1.053695(63) \\
\hline \hline
\end{tabular}
\end{center}
\end{table}

We analyzed the series by the numerical method of differential approximants 
\cite{Guttmann89}. Estimates of the critical point and critical exponents 
were obtained by averaging values obtained from inhomogeneous differential 
approximants chosen such that most, if not all, series
terms were used. Some approximants were excluded from the averages because 
the estimates were obviously spurious. The error quoted for these estimates 
reflects the spread (basically one standard deviation) among the approximants. 
Note that these error bounds should {\em not} be viewed as a measure of the 
true error as they cannot include possible systematic sources of error. In 
Table~\ref{tab:analysis} we have listed the results from this analysis. 
Our first observation is that the critical points of open and semi-meanders
are identical, and thus so are the connective constants $R$ for all
the problems (recall that for closed meanders $x_c=1/R^2$). Clearly
the most accurate estimates are obtained from the semi-meander series
and from this we estimate, conservatively, that $x_c=0.2855644(2)$ and
thus $R=3.501837(3)$. Secondly, as expected open and closed meanders
have the same critical exponent, which we estimate to be 
$\alpha=3.4208(6)$. This could, though only marginally, be consistent with
the conjectured value $\alpha= (29+\sqrt{145})/12 = 3.4201328\ldots$. 
For semi-meanders we estimate $\overline{\alpha}=2.0537(2)$, which
is not consistent with the conjecture 
$\overline{\alpha}= 1+\sqrt{11}(\sqrt{29}+\sqrt{5})/24 = 2.0531987\ldots$.

\begin{figure}
\includegraphics{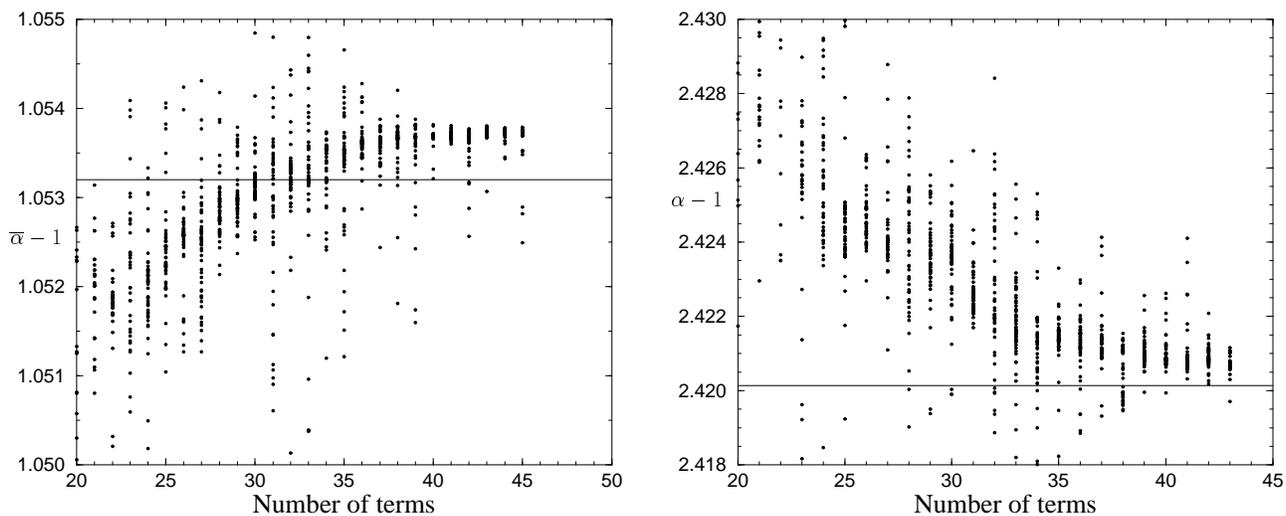}
\caption{\label{fig:ntexp} Estimates of the critical exponents of
the semi-meander generating function, $\overline{\alpha}-1$, and 
the open meander generating function, $\alpha-1$, vs. the number of 
terms from the series used by the differential approximants. Each
point represents an estimate obtained from a particular 2. order
differential approximant.}
\end{figure}

\begin{figure}
\includegraphics{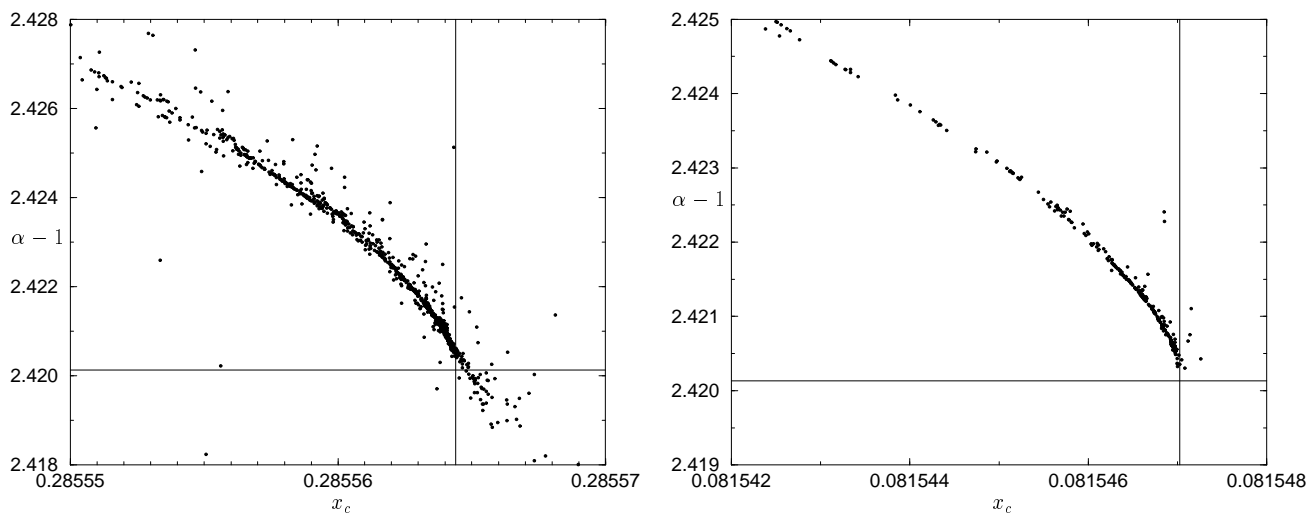}
\caption{\label{fig:crpexp} Estimates of the critical exponents for open 
(left panel) and closed meanders vs. the corresponding critical point 
estimates.}
\end{figure}

In order to gain a better understanding of the behaviour of the
exponent estimates it is useful to plot them against the number of
terms used to form the differential approximant. In particular we can 
check whether or not the estimates asymptote or whether they are
drifting with the length of the series. 
In figure~\ref{fig:ntexp} we have done this for semi-meanders and
open meanders. These plots strongly reinforce the comments made
above. The exponent estimates for semi-meanders increase as 
more terms are used and appear to settle down to an asymptotic
value above the conjectured value.  
For open meanders the exponent
estimates decrease and approach the conjectured value as more terms
are used. It is quite likely that with a longer series the estimates
would actually converge to the conjectured value, though it is also
possible that the estimates could settle at a value just above
the conjectured value. If we look at the estimates in 
Table~\ref{tab:analysis} we note that for both open and closed
meanders the exponent estimates decrease as the critical point
estimates increase. It is possible that as $x_c$ approaches
the estimate obtained from semi-meanders the exponent estimates
approach the conjectured value. To check this we plotted (in
figure~\ref{fig:crpexp}) the exponent estimates
vs the critical point estimates for open and closed meanders. The
solid lines are the conjectured exponent value and the best
estimate for the critical point based on the semi-meander analysis.
Clearly the estimates pass extremely close to the intersection between
the solid lines, lending further support to the possibility
that the conjecture for $\alpha$ is correct.

Next we looked for non-physical singularities and found that
both the open and semi-meander generating functions have a
singularity at $-1/R$ with an exponent whose value is
consistent with $\alpha-1$. These generating functions also
have a pair of singularities in the complex plane at 
$\pm 0.685(5)i$. The exponent estimates are quite poor, but
consistent with the value $\alpha-1$.

Finally we turned our attention to the ``fine-structure'' of the
asymptotic behaviour of the meandric numbers,

\begin{eqnarray}
M_n &\sim & R^{2n}\sum_{i=0} c_i/n^{\alpha+f(i)}, \\
m_n &\sim & R^{n}\sum_{i=0} [c_i/n^{\alpha+f(i)}+(-1)^nd_i/n^{\alpha+f(i)}], \\
\overline{M}_n &\sim & R^{n}\sum_{i=0} [c_i/n^{\overline{\alpha}+f(i)}
+(-1)^nd_i/n^{\alpha+f(i)}].
\end{eqnarray}
\noindent
The alternating sign terms are due to the singularity at $-1/R$.
Fitting the meandric numbers to these formulas we found excellent
convergence when $f(i)=i$. This corresponds to the case where
there are only analytic corrections-to-scaling terms. 
The leading amplitudes $c_0$ are of special interest, and 
for closed, open and semi-meanders we found the values
0.339(1), 11.45(3), and 0.688(1), respectively.

\section*{E-mail or WWW retrieval of series}

The series for the various generating functions studied in this paper
can be obtained via e-mail by sending a request to 
I.Jensen@ms.unimelb.edu.au or via the world wide web on the URL
http://www.ms.unimelb.edu.au/\~{ }iwan/ by following the instructions.

\section*{Acknowledgements}

We would like to thank Di Francesco et al. for sending us their papers 
prior to publication.
Financial support from the Australian Research Council is gratefully 
acknowledged.


\begin{thebibliography}{10}

\bibitem{LZ} S. K. Lando and A. K. Zvonkin, 
Theoret. Comput. Sci. {\bf 117}, 227 (1993).

\bibitem{Poincare} H. Poincar\'e, Rend. Circ. Mat. Palermo {\bf 33},
375 (1912).

\bibitem{Touchard} J. Touchard, Canad. J. Math. {\bf 2}, 385 (1950).

\bibitem{Koehler} J. E. Koehler, J. Combin. Theory {\bf 5}, 135 (1968).

\bibitem{Lunnon} W. Lunnon, Math. Comp. {\bf 22}, 193 (1968).

\bibitem{Arnold} V. Arnold, Siberian Math. J., {\bf 29} 717 (1988).

\bibitem{KS} K. H. Ko and L. Smolinsky, 
Pacific J. Math. {\bf 149}, 319 (1991).

\bibitem{HMRT} K. Hoffmann, K. Mehlhorn, P. Rosenstiehl, and R. E. Tarjan,
Information and Control {\bf 68}, 170 (1988).

\bibitem{AM} N. Alon and W. Maass, 
J. Comput. System Sci. {\bf 37}, 118 (1988).

\bibitem{FGG1} P. Di Francesco, O. Golinelli and E. Guitter,
Math. Comput. Modelling {\bf 26}, 97 (1997).

\bibitem{FGG2} P. Di Francesco, O. Golinelli and E. Guitter,
Nucl. Phys. B {\bf 482}, 497 (1996).

\bibitem{FGG3} P. Di Francesco, O. Golinelli and E. Guitter,
Commun. Math. Phys. {\bf 186}, 1 (1997).

\bibitem{Francesco}  P. Di Francesco, 
Commun. Math. Phys. {\bf 191}, 543 (1998);
J. Math. Phys. {\bf 38}, 5905 (1997).

\bibitem{Makeenko} Y. Makeenko, 
Nucl. Phys. Proc. Suppl. {\bf 49}, 226 (1996).

\bibitem{SS} G. W. Semenoff and R. J. Szabo, 
Int. J. Mod. Phys. {\bf A12}, 2135 (1997).

\bibitem{FGG4} P. Di Francesco, O. Golinelli and E. Guitter,
Nucl. Phys. B {\bf 570}, 699 (2000).

\bibitem{FGJ2000} P. Di Francesco,  E. Guitter and J. L. Jacobsen,
preprint http:://arxiv.org/abs/cond-mat/0003008.

\bibitem{Jensen99} I. Jensen,  
preprint http:://arxiv.org/abs/cond-mat/9910313

\bibitem{Ent} I. G. Enting, J. Phys. A {\bf 13}, 3713 (1980).

\bibitem{Guttmann89} A. J. Guttmann, in {\em Phase Transitions and
Critical Phenomena}, Vol. 13, eds. C. Domb and J. L. Lebowitz,
Academic Press, New York (1989), pp 1-234.


\end{thebibliography}
\end{document}